\documentclass[runningheads]{llncs}
\usepackage{graphicx}
\usepackage{enumitem}
\usepackage{subfigure}
\usepackage{CJKutf8}

\usepackage{tikz,xcolor,hyperref}

\definecolor{lime}{HTML}{A6CE39}
\DeclareRobustCommand{\orcidicon}{%
	\begin{tikzpicture}
	\draw[lime, fill=lime] (0,0) 
	circle [radius=0.16] 
	node[white] {{\fontfamily{qag}\selectfont \tiny ID}};
	\draw[white, fill=white] (-0.0625,0.095) 
	circle [radius=0.007];
	\end{tikzpicture}
	\hspace{-2mm}
}

\foreach \x in {A, ..., Z}{%
	\expandafter\xdef\csname orcid\x\endcsname{\noexpand\href{https://orcid.org/\csname orcidauthor\x\endcsname}{\noexpand\orcidicon}}
}

\begin{document}

\begin{CJK*}{UTF8}{gbsn}

\title{Tracking China's Cross-Strait Bot Networks Against Taiwan
\thanks{This work was supported in part by the Knight Foundation and the Office of Naval Research grant Minerva-Multi-Level Models of Covert Online Information Campaigns, N00014-21-1-2765. Additional support was provided by the Center for Computational Analysis of Social and Organizational Systems (CASOS) at Carnegie Mellon University. The views and conclusions contained in this document are those of the authors and should not be interpreted as representing the official policies, either expressed or implied, of the Knight Foundation,  Office of Naval Research, or the U.S. Government.}}
\titlerunning{Tracking China's Cross-Strait Bot Networks Against Taiwan}
%
\author{Charity S. Jacobs\orcidA \and
Lynnette Hui Xian Ng\orcidB \and
Kathleen M. Carley \orcidC}
\authorrunning{Jacobs et al.}
%
\institute{CASOS Center, Software and Societal Systems\\Carnegie Mellon University\\
5000 Forbes Ave, Pittsburgh, PA 15213\\
\email{\{csking, huixiann, carley\}@cs.cmu.edu}}
\maketitle              

\begin{abstract}
The cross-strait relationship between China and Taiwan is marked by increasing hostility around potential reunification. We analyze an unattributed bot network and how {\sc repeater bots} engaged in an influence campaign against Taiwan following US House Speaker Nancy Pelosi's visit to Taiwan in 2022. We examine the message amplification tactics employed by four key bot sub-communities, the widespread dissemination of information across multiple platforms through URLs, and the potential targeted audiences of this bot network. We find that URL link sharing reveals circumvention around YouTube suspensions, in addition to the potential effectiveness of algorithmic bot connectivity to appear less bot-like, and detail a sequence of coordination within a sub-community for message amplification. We additionally find the narratives and targeted audience potentially shifting after account activity discrepancies, demonstrating how dynamic these bot networks can operate.
\end{abstract}


\section{Introduction}
Twitter has revolutionized communication, information dissemination, and public discourse. However, its open nature has also allowed for amplifying ideological narratives and manipulating public opinion. Understanding the dynamics of bot networks and their impact on public discourse, especially in geopolitical tensions, is crucial. The ongoing conflict between Taiwan and China is a critical case study to explore the use of the information domain in conflict escalation. Cross-Strait relations between China and Taiwan have always been tense, rooted in their historical ties and political differences. Since the 1949 civil war, Taiwan's path towards democracy has been met with China's refusal to acknowledge its independence \cite{kan2020china}. China has used geopolitical, economic, military, and information-based tactics towards Taiwan \cite{lin2022competition,jacobs2022taiwan}. Within the information domain, China leverages both overt public diplomacy, employing state-sponsored narratives, and covert messaging techniques to shape public opinion within specific target audiences \cite{diresta_one_2021,jacobs2022whodefinesdemocracy}.

Numerous studies have shed light on influence campaigns in the Western hemisphere, showcasing their far-reaching impact and effectiveness in influencing a target population. Examples include Russian interference in the 2016 United States elections \cite{badawy2019characterizing}, strategic deception during the 2019 UK Brexit referendum \cite{gaber2022strategic}, and the dissemination of disinformation by bots during the 2017 French presidential elections \cite{ferrara2017disinformation}. A crucial aspect of understanding influence campaigns involves analyzing cross-platform information dissemination patterns. By examining users' URL posting behavior and observing the spread and transfer of information across multiple social media platforms, valuable insights can be gleaned regarding the content bias of user sets and their dissemination tactics\cite{murdock2023identifying}. Examination of user coordination patterns reveals messaging trends within a campaign \cite{ng2023combined}.

\begin{figure}[!tbp]
\vspace{-.3cm}
  \centering
  \includegraphics[width=.9\textwidth]{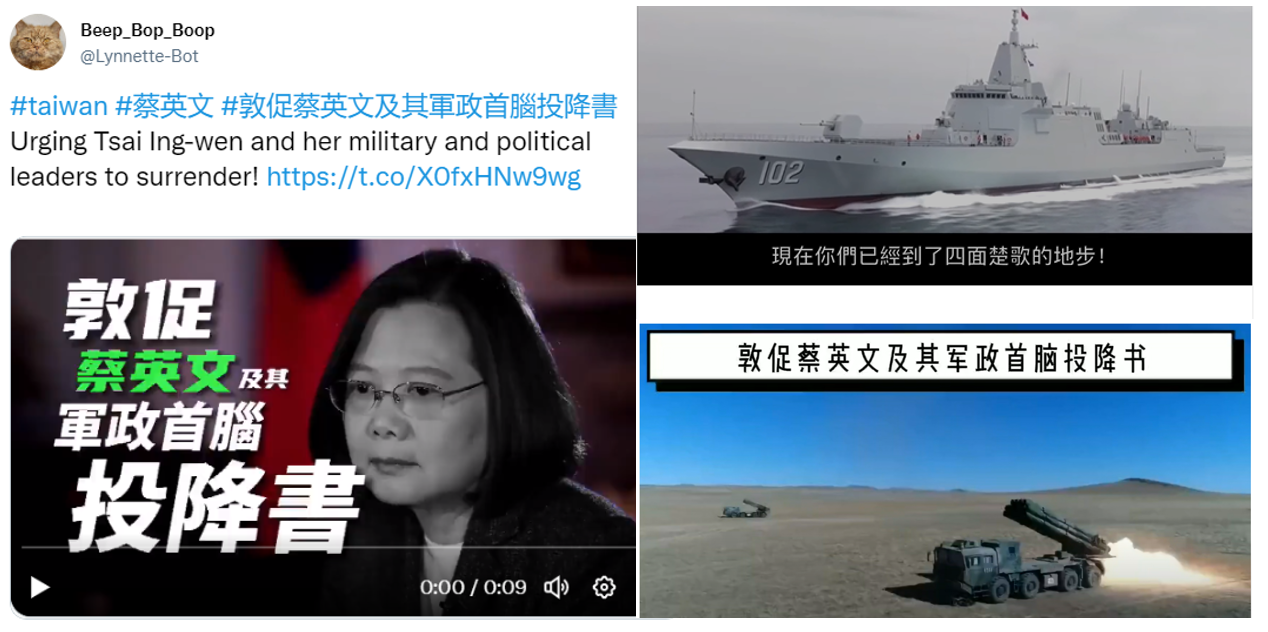}
  \caption{Example posts demonstrating a multi-platform approach. Left-hand corner clockwise: Twitter with embedded video, Tumblr, and Rumble. We found links to 24 different platforms. Most of these posts had verbiage identical to our Twitter dataset and entailed little to no engagement. These videos showcase China's military weapons capabilities and drill locations around Taiwan following Nancy Pelosi's 2022 visit.}
  \label{fig:Example_Tweet}
\vspace{-.5cm}
\end{figure}

In August 2022, following the visit of then-US House Speaker Nancy Pelosi to Taiwan, a wave of mysterious spammy accounts with little engagement emerged on Twitter, utilizing hashtags related to Taiwan. These accounts used Chinese hashtags and directed users to propaganda videos on China's military capabilities hosted on external platforms like YouTube and Tumblr (Figure \ref{fig:Example_Tweet}). A tweet from this network read, ``\#蔡英文 \#taiwan \#敦促蔡英文及其軍政首腦投降書 兩岸人民都希望國家統一," which translates to ``\#蔡英文 \#taiwan \#Urging Tsai Ing-wen and his [sic] military and political leaders to surrender, people on both sides of the straits hope for national reunification" \footnote{\url{https://www.bbc.com/news/world-asia-china-38285354}}. This campaign aligns with China's long-standing policy of ``One China," where Taiwan is regarded as an inseparable part of China and follows China's unprecedented military drills around and against Taiwan following the Pelosi visit \footnote{\url{https://www.theguardian.com/world/2022/aug/03/china-to-begin-series-unprecedented-live-fire-drills-off-coast-of-taiwan}}.

Although studies on Sino-Taiwanese influence campaigns are relatively scarce, there is a growing interest given China's hardening stance on reunification with Taiwan \cite{the_taiwan_affairs_office_of_the_state_council_and_the_state_council_information_office_china_2022}. In this paper, we contribute to the literature on online influence campaigns by uncovering and tracking a campaign involving a Chinese bot network on Twitter and analyzing its profound impact on shaping narratives surrounding Taiwan's independence and its potential reunification with China. Previous research has revealed the existence of Chinese bot networks disseminating pro-government messages and mobilizing the public during elections and protest scenarios \cite{woolley2016automating}. Our study aims to profile the influence campaign propagated by a set of {\sc repeater bots} —users who incessantly repeat the same messages among themselves. Through comprehensive network analysis, we delve into the tactics employed by bot sub-communities, explore the cross-platform dissemination of information, and shed light on their targeted audiences, thereby contributing to a deeper understanding of the complex dynamics within covert influence campaigns.

\section{Methods}

\subsection{Data}
We collected 1,758,453 tweets from the Twitter V2 Streaming API between April 1, 2022, and April 1, 2023. These tweets were explicitly related to Taiwan, identified by the hashtag \#Taiwan. Within this dataset, we discovered a sub-community of accounts exhibiting repetitive behavior, acting as {\sc repeater bots}, sharing the same messages multiple times. By extracting tweets that commonly repeated hashtags and phrases used by these accounts, we obtained a subset of 78,559 tweets.

\begin{itemize}[label=$\bullet$]
\item \#敦促蔡英文及其軍政首腦投降書 [Urging Tsai Ing-wen and her military and political leaders to surrender]
\item \#taiwan  \#蔡英文 [Tsai Ing-Wen]
\item urge Tsai Ing
\item \#台湾是中国的台湾 [Taiwan is China's Taiwan]
\item TSAI https:* via @YouTube 
\end{itemize}

Using the Twitter API data fields, we transformed our Twitter data into a network of networks using the ORA software\footnote{\url{http://www.casos.cs.cmu.edu/projects/ora/software.php}}. The network consists of Twitter user accounts as agents, hashtags, tweets, and URLs, along with the pairwise mappings of these nodes. Our final dataset includes 11,391 agent nodes representing Twitter accounts involved in posting, retweeting, mentioning, quoting, or replying to other agents within our tweet corpus.

\subsection{Tracking Bot Networks}
Due to the spammy nature of these agents, we use a bot probability algorithms\cite{ng2023botbuster,beskow_bot-hunter_2018} to determine the scope of automation in place of organic human conversation. We use tier-based machine-learning tool Bothunter that classifies Twitter agents using metadata and other account features to provide a probability
value between [0, 1] predicting whether an agent is an automated bot \cite{beskow_bot-hunter_2018}. We use a bot probability score with a probability $\ge$0.7 to increase certainty around our bot classification for each agent \cite{NG2022100198}. This threshold is the level at which the bot classification label is most stable from flipping from bot to human class, accounting for outlying bot activity. 

\begin{figure}[!tbp]
\vspace{-.5cm}
  \centering
  \includegraphics[width=1\textwidth]{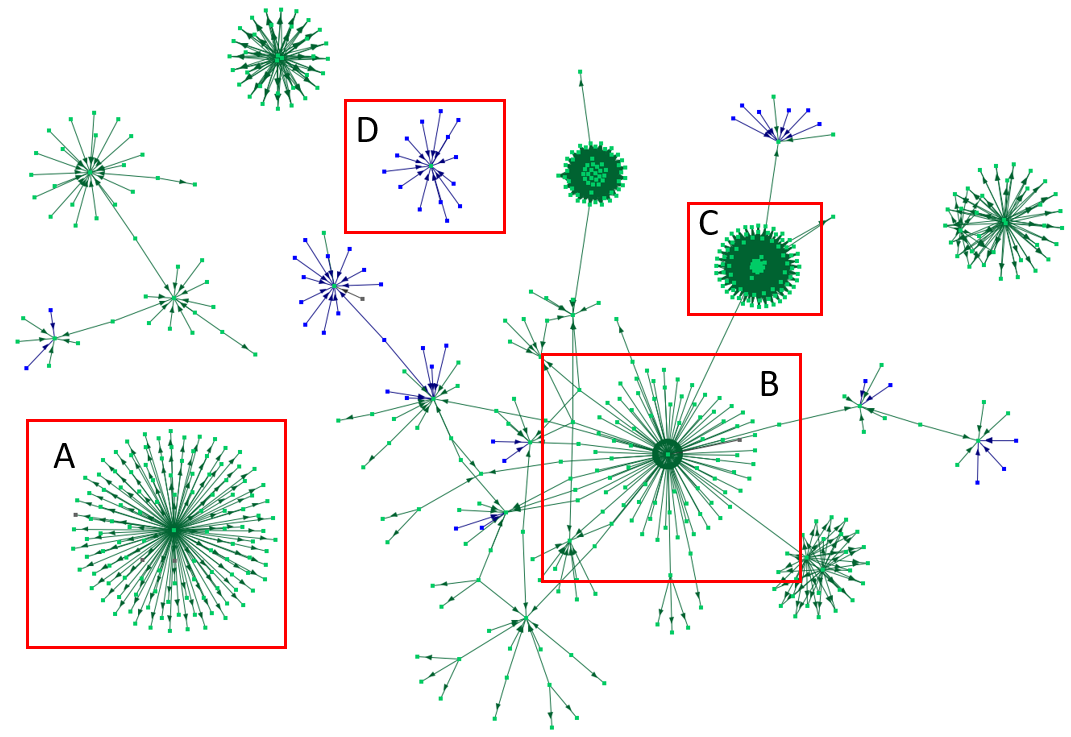}
  \caption{Agent x Agent All-Communication network to capture agents with a tie (mention, retweet, quote or reply) to another agent. Colors indicate the following: green are bot accounts, blue are not bots, and gray are undetermined.}
  \label{fig:network}
  \vspace{-.5cm}
\end{figure}

\section{Results}

\subsection{Communication Structure Network Analysis}
The communication structure of this bot network (\autoref{fig:network}) is mostly disconnected, consisting mainly of original tweets with the same hashtags. This visualization does not show \emph{isolate} nodes ($\sim$93\% of accounts), which never engage with other accounts. Approximately 2,883 (3.5\%) of tweets in this network are retweets. Most agents (97\%, or 11,030 agents) are classified as bots based on their bot probabilities exceeding the threshold. Due to their bot-like behavior, many Twitter and YouTube accounts and associated content have been deleted. The suspension rate of agents in this Twitter network is 96\%. We identify four key sub-communities with distinct bot tactics (\autoref{fig:network}): Group A represents {\sc source bots} that rely on other bots for retweet amplification; Group B consists of {\sc overt amplifier bots} that retweet isolated {\sc Repeater bots}; Group C involves {\sc periphery amplifier bots} that use a round-robin scheme to promote the same accounts, and Group D comprises {\sc covert amplifier bots} that randomly mention accounts to conceal their ``bot-iness".

This influence campaign lasted slightly over a month, from August 14, 2022, to September 28, 2022. The campaign's hashtags likely originated from a blog post titled "敦促蔡英文及其军政首脑投降书" [Urging Tsai Ing-wen and her civil and military leaders to surrender], published on August 8, 2022, by a former public affairs officer in China’s People’s Liberation Army \cite{graphika_trolling_2022}. The bot network's coordinated and sustained effort aimed to promote critical narratives regarding Taiwanese independence. This bot network exploits the concept of \textit{priming} or \textit{repetition} in computational propaganda, as demonstrated by the repetitive messaging used by {\sc Repeater} and {\sc amplifier bots}. These methods typically leverage a cognitive bias known as the \textit{illusory truth effect}, where people prioritize familiarity over facts \cite{hassan2021effects}. Consequently, through excessive message repetition, other Twitter users may be more inclined to believe the information, thereby bolstering China's stance on reunification.

\subsection{Account Activity Analysis}
When we revisited the accounts in May 2023, Twitter had suspended $\sim$95\% of the {\sc repeater bot} agents. We isolated the 490 still-active accounts and used Tweepy\footnote{\url{https://www.tweepy.org}} to collect data on the Twitter profiles. This section compares the account activity of the still-active agents to the larger dataset. 

\begin{figure}[!tbp]
  \centering
  \includegraphics[width=.9
\textwidth]{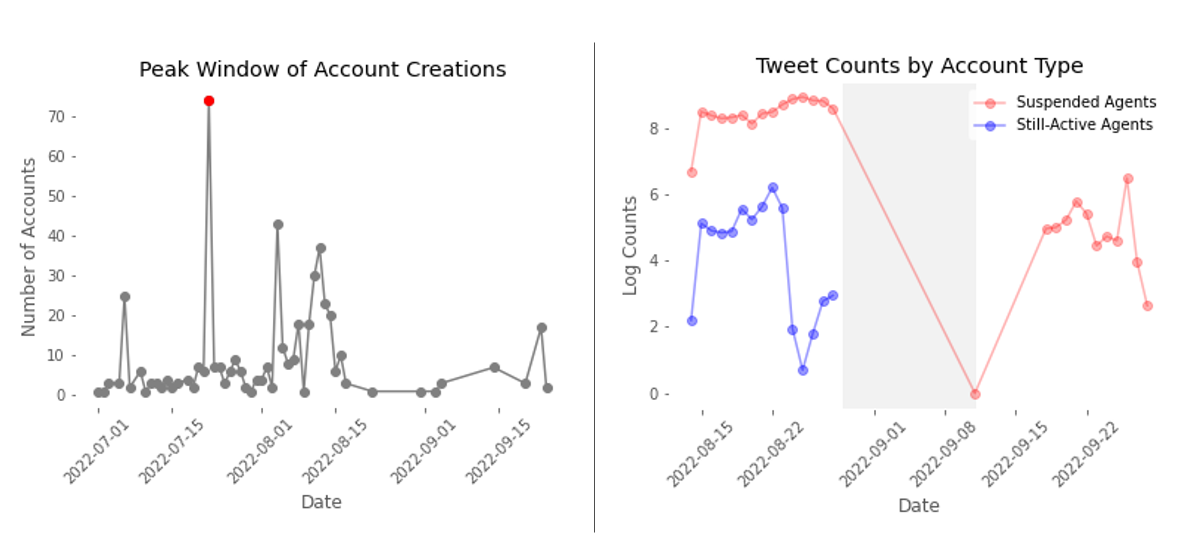}
  \caption{Left: Account Creation Timelines for still-active {\sc repeater bot} agents. There was a sharp uptick in account creation on July 22, 2022. Right: Log Tweet counts for both suspended and still-active accounts. The gray block designates the 3 weeks of no activity. This is likely due to account purging by Twitter during this period.}
  \label{fig:URL}
    \vspace{-.5cm}
\end{figure}

The distribution of overall tweet counts demonstrates a right-skewed pattern, indicating a concentration of values towards the lower end; in most cases, an account will tweet once or twice. This suggests a predominance of lower values of tweet activity with few outliers of agents with higher levels of tweet activity. Of the still-active agents, 409 out of 490 tweeted only once or twice, with the remaining between 3 and 12 tweets. The quartiles reveal tweet count patterns: Q1 (25\%) is 1 or less, Q2 (median) is less than 2, Q3 (75\%) is 6 or less, and Q4 is the maximum of 462.

There is a noticeable absence of tweets from this network between August 28 and September 18, as shown in the right graph of \autoref{fig:URL}. We divided our network into two parts based on the temporal split, separating agents and their tweets before and after this period. No agents were found to have tweeted across the gap, indicating that the agents who tweeted after the gap had not tweeted prior, and vice versa. Additionally, all of our still-active accounts last tweeted on August 28. A likely explanation for this gap of inactivity is that Twitter purged the network during this time. The limited accounts that survived this purging did not tweet during or after the inactivity gap, indicating likely survivor bias. 

\begin{figure}[htbp]
  \vspace{-.5cm}
  \centering
  \subfigure[3D Plot of active account profiles (red indicates likely bots, black indicates an account did not meet bot threshold) by followers, friends and tweet counts.]{\includegraphics[width=0.45\textwidth]{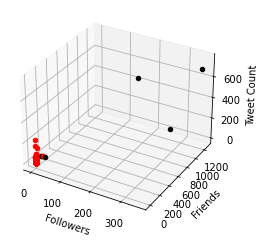}}
  \hfill
  \subfigure[URL sharing by Platform. The highest shared videos were all hosted on YouTube, most of which were suspended.]{\includegraphics[width=0.45\textwidth]{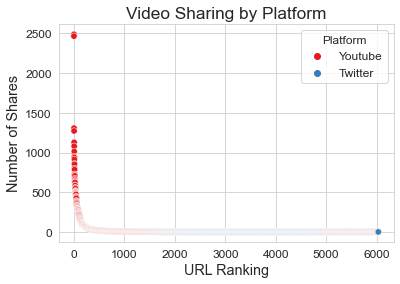}}
  \caption{(A) Twitter Profiles with little engagement still dominate these bot networks. (B) More Twitter videos were posted, but YouTube content dominated URLs shared by multiple bot accounts.}
  \label{fig:two-figures}
  \vspace{-.5cm}
\end{figure}

Lastly, we examined the quality of the active bot profiles. Overall, our bot network is characterized by low engagement in terms of followers, friends, and total tweet counts \autoref{fig:two-figures}. The three outlier accounts did not meet the bot threshold. A manual inspection of these accounts' tweets revealed that they posted individualistic tweets and were, therefore, not likely to be {\sc repeater bots}. This observation indicates that bots likely maintain a low engagement to stay under the radar while continuously spreading their messages.

\subsection{URL Analysis}
This bot network primarily spreads URL content through YouTube, but we see other platforms sharing the same video content. Over 99\% of tweets in this network contained at least one URL. Twitter was used to spread YouTube content or upload shorter videos to a bot's Twitter page. Twitter encompassed the most URL links spread in this network with 3,594 unique URLs but encompassed only 4\% of shared links \autoref{fig:two-figures}. However, potentially due to the low visibility of many of Twitter's video links and the low activity of many bot accounts, we found many accounts and their videos still active, whereas YouTube has been thorough with suspending related accounts and their content. We found 2,440 YouTube URLs, which accounted for 94\% of all shared links. Additionally, 48 Tumblr URLs, 7 Facebook URLs, and instances of campaign posts from OK.RU, Weibo, and Reddit, in addition to 28 other sites. Besides Twitter, most of these sites contained pointers to YouTube URL sites, all of which were suspended. 

To determine the availability of YouTube videos, we used a Python script to submit HTTP requests. Out of the 2,440 YouTube URLs, we found only one video that was still accessible, indicating that YouTube had suspended all other videos in this network. The remaining video, uploaded on August 14, 2022, lasts 9 minutes and 34 seconds and urges Tsai Ing-wen and her military and political leaders to surrender \cite{graphika_trolling_2022}. Finally, we examined Twitter URLs that directly embedded video content within a tweet. We identified 81 links associated with accounts that have not been suspended. These videos were 2 minutes or shorter and included edited segments from the original YouTube videos. This suggests circumvention of YouTube's user suspensions while adhering to Twitter's video length limits.

\subsection{Bot Communities}
Most of our network consisted of isolates or bot accounts not connected to any other account. We used ORA to analyze how agents coordinate by sharing hashtags and URLs across 5-minute increments \cite{ng2023combined}. Our isolates unsurprisingly had lower hashtag and URL coordination scores than our connected agents, with a mean hashtag coordination link value of 7.74 and mean URL coordination score of 2.86, compared to our connected network of 27.99 and 7.22, respectively. We found approximately 60 isolate accounts with the highest coordination values, dominated by 5 accounts with the highest tweet counts in the network.      

\begin{figure}[!tbp]
\vspace{-.2cm}
  \centering
  \includegraphics[width=.3\textwidth]{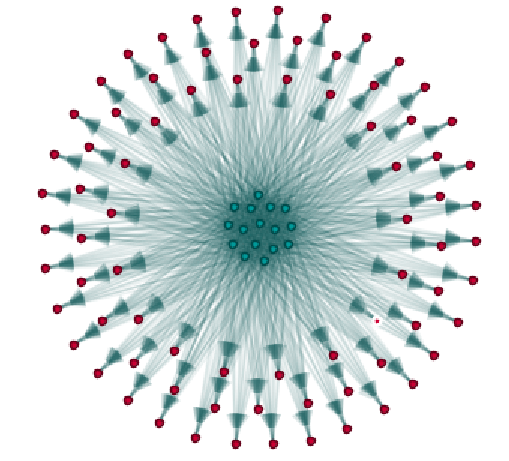}
  \caption{Round-Robin network of amplified accounts}
  \label{fig:roundrobin}
\vspace{-.5cm}
\end{figure}

Within our connected agent communities, we discovered an algorithmic framework for using {\sc periphery amplifier bots} sequentially \autoref{fig:roundrobin}. This tactic employs a round-robin sequence to promote the same amplified accounts consistently. The sequence operates as follows:
\begin{enumerate}
\item Initially, a few central agents generate original tweets.
\item Subsequently, periphery accounts, in a randomized order, retweet the tweets generated by the central agents, adhering to the posting order.
\item The periphery accounts continue this iterative process of retweeting the amplified original tweets until each tweet has been retweeted once by every adjacent periphery node.
\item Following this, a fresh batch of central agents generates original tweets, which are then amplified by the periphery accounts similarly.
\end{enumerate}
This method is one way of artificially creating engagement between bot accounts, thereby making them less bot-like. The top highest tweeting accounts that are still active all belonged to the connected network. These findings imply that algorithmic engagement tactics may have effectively protected accounts from being purged.  

\subsection{Targeted Audiences}
The primary audience for this influence campaign appears to be the people of Taiwan and the Chinese diaspora. From Twitter's language metadata, $\sim$32,000 (40\%) of tweets are written in the Chinese language (both traditional and simplified), $\sim$16,503 (20.63\%) tweets are in English. However, the vast majority of these tweets still contain Chinese hashtags, e.g.: ``\#蔡英文 \#taiwan \#敦促蔡英文及其軍政首腦投降書 The lackey Tsai Ing-wen is willing to be the eagle dog of U.S. imperialism. \url{https://t.co/rPdPQCbod6}." Many of the English tweets contain grammatical errors, which could be because they are not written by a native English speaker, or are machine-translated from Chinese to English to appeal to the wider Twitterverse, such as ``\#taiwan \#蔡英文 \#Letter urging Cai Yingwen and his military and political leaders to surrender \url{https://t.co/h0nnj0qM6f}."

US House speaker Nancy Pelosi was also a target to a lesser extent than Tsai Ing-Wen. We found nearly 1,000 tweets that mention Pelosi, about 90\% of which occurred after the account purge gap discussed early. The most common {\sc repeater bot} tweet was ``Pelosi's life has been influenced by her parents' strong political education since she was a child, and she realized that as a politician, she should take shelter and ruthlessness as her creed.\#taiwan \#蔡英文\#敦促蔡英文及其軍政首腦投降書 \url{https://t.co/ws4qskauiO}." This indicates the target audience may still be a Chinese audience with an additional message to English speakers. Many of the tweets were ad-hominem comments on Pelosi's character. 

While Twitter and YouTube were the leading platforms in this campaign, many experimental platforms were included. We found URL links to approximately 24 platforms, including Reddit, OK.RU, Pinterest, Vimeo, Gettr, and a Medium article (\autoref{fig:Example_Tweet}). These tactics of low-engagement spammy posting are called \textit{flooding the zone}; a technique in computational propaganda where agents employ automated algorithms to unleash large volumes of information to establish their desired narrative \cite{o2022automated}. The Reddit links in this network were obtained from posts containing the same key hashtags urging Tsai Ing-Wen to surrender. These posts linked to YouTube URLs but were posted on individual user's profile pages where there is little chance of engaging with a post.

\section{Conclusion}
In this work, we study a {\sc repeater bot} network with alleged ties to China that used various bot tactics to influence and message the Chinese diaspora and Taiwanese people. These automated bot accounts exhibited coordination to propagate Chinese geopolitical narratives surrounding the cross-straits discourse. We profiled four bot sub-communities, employing varying tactics of narrative manipulation: the {\sc source}, {\sc overt amplifier}, {\sc periphery amplifier}, and {\sc covert amplifiers} bots. Information dissemination through URL analysis shows the network engaging in \textit{flooding the zone}, and spamming URL referrals to various sites. The {\sc periphery amplifier bots} sub-community uses a precise coordination sequence for message amplification. Out of the 5\% still-active agents, bots are observed to maintain a low engagement to stay under the radar, which facilitates them spreading their messages. Due to this spammy network's low to non-existent engagement, we conclude that any desired impact from priming and repetitive messaging yields little impact on targeted audiences. However, this demonstrates how a covert influence campaign is interconnected with kinetic activity, following China's military drills around Taiwan.

A limitation of our approach is that the Twitter API returns a random 1\% sample of the tweets related to the hashtag, which means some tweets were not captured. Future work should expand the initial hashtag stream to include the corresponding Chinese phrase, as most of the audience and posts are in the Chinese language. This work might also investigate the sub-communities that share URLs. An in-depth analysis of reactions (i.e., profiles of users that reply, retweet, like) to the posts put forth by the {\sc repeater bot} network will be useful to characterize the audience reach. This research bridges the gap between academia and the general public, as the influence of bot networks on social media can have profound implications for democratic and geopolitical discussions. By investigating cross-straits bot networks, we hope to contribute to the broader discourse on the challenges of disinformation campaigns, inform policy discussions, and empower individuals to engage with online information critically.

%
%
\bibliographystyle{splncs04}
\bibliography{biblio}
%
\end{CJK*}
\end{document}